\documentclass[showpacs,aps,pra,twocolumn,amsmath,amssymb]{revtex4-1}
\usepackage{graphicx}
\usepackage{dcolumn}
\usepackage{bm}
\usepackage{epsf}
\usepackage{amsmath}
\usepackage{color}

\newcommand{\ket}[1]{|#1\rangle}
\usepackage{bm}
\usepackage{lettrine}
\usepackage{dcolumn}
\usepackage{bm}
\usepackage{graphicx}
\usepackage{amsmath}
\usepackage{latexsym}
\usepackage{amsfonts}
\usepackage{amssymb}
\usepackage{array}
\usepackage{epsfig}
\usepackage{epstopdf}
\usepackage{times}
\usepackage{amsmath,epsfig}
\usepackage{tikz}

\begin{document}

\title{Entangling power of two-qubit gates on mixed states}

\author{Zhe~Guan$^1$}
\author{Huan~He$^1$}
\author{Yong-Jian~Han$^1$}
\author{Chuan-Feng~Li$^1$}\email{cfli@ustc.edu.cn}
\author{Fernando~Galve$^2$}\email{fernando@ifisc.uib-csic.es}
\author{Guang-Can~Guo$^1$}

\affiliation{$^1$ Key Laboratory of Quantum Information, Chinese Academy of
Sciences,
University of Science and Technology of China, Hefei 230026, People's Republic
of
China}

\affiliation{$^2$ IFISC (UIB-CSIC), Instituto de F\'{i}sica Interdisciplinar y
Sistemas Complejos,
UIB Campus, E-07122 Palma de Mallorca, Spain}

\begin{abstract}
The ability to reach a maximally entangled state from a separable one through
the use of a two-qubit unitary operator is analyzed for mixed states. This extension
from the known case of pure states shows that there are at least two families of gates
which are able to give maximum entangling power for all values of purity. It is notable
that one of this gates coincides with a maximum discording one. We give analytical
proof that such gate is indeed perfect entangler at all purities and give numerical
evidence for the existence of the second one. Further, we find that there 
are other gates, many in fact, which are perfect entanglers for a restricted range of purities.
This highlights the fact that many perfect entangler gates could in principle be found
if a thorough analysis of the full parameter space is performed.
\end{abstract}


\maketitle

\section{Introduction}
Entanglement is a nonlocal resource which has been widely
investigated~\cite{Ho09} and applied to quantum information tasks such as
quantum communication ~\cite{Sc96}, quantum computation~\cite{Dp95,V98,N00}, and
quantum teleportation~\cite{Pan01}. For many years it was thought to be the main 
source of quantum advantage with respect to classical information and computation
tasks. However, in the recent years a new figure of merit for quantumness, the quantum
discord \cite{discord} has attracted much attention as another possible,
inequivalent, source of advantage (see \cite{reviewVedral} for a review on this measure).
Because it is not anymore clear where the source of advantage lies, it is of theoretical
interest to check what could be the main differences between these two measures
and their application to quantum information tasks. Since they coincide for pure
states, the crux of this situation might revolve around mixed states, and therefore
it is of crucial interest to move to the realm of mixed states.

The problem of finding the unitary operators which are able to
produce maximum entanglement from two-qubit separable pure states 
was intensively studied some years ago by several authors (see for example \cite{JZ,Linden,Bala,C96,BKJ01,M01,zanardi,zanardi2}),
but the extension to mixed states has remained largely unexplored \cite{plastino}. Recently, the generation 
of maximum quantum discord by two-qubit gates was studied in \cite{fer13}, where they found
one family of gates able to generate maximum discord from classical-classical states (i.e. states $\rho_{cc.}^\mu$
with purity $\mu=Tr[(\rho_{cc.}^\mu)^2]$) valid for all purities (i.e. all possible values of mixedness).
This family included the $\sqrt{\rm SWAP}$ gate.
In addition, it was found that other gates could produce maximum discord for whole ranges of
purities, such as for example the CNOT gate which is perfect discorder for all purities except for 
the range of rank 2 states. With these intriguing results in mind, one might wonder whether those gates
are also perfect entanglers and why. We set out to give a first step in answering this question.

In this work we will show analytically that the perfect discorder gate found in \cite{fer13} is
also a perfect entangler for all purities. Further, we will also show that the gates which were
found to be perfect discorders only for partial range of purity in \cite{fer13}, are still
perfect entanglers for the same range. This is an amazing coincidence, considering that
entanglement and discord are two measures with a very different definition, and
is worthy of further work, since probably some fundamental insight can be gained from its study.

We will also provide strong numerical evidence that the family of perfect entangler gates to which
the CNOT gate belongs, are indeed so for all range of purities. In addition, we will discover many
more families which are able to reach maximum entanglement for partial ranges of purity, something
which highlights the need for a more thorough investigation.

\section{Entangling power of a unitary}

Despite several definitions have been used in the literature, we will focus in the following:

\begin{equation}
EP_\mu(U)=\max_{\rho_{\rm sep}^\mu}E_F[U\rho_{\rm sep}^\mu U^\dagger]
\end{equation}
where $\rho_{\rm sep}^\mu$ is a separable state of purity $\mu$ (i.e. tr$[(\rho_{\rm sep}^\mu)^2]=\mu$) and
$E_F(\rho)$ is the entanglement of formation of $\rho$:
\begin{equation}
E_F(\rho)=\min_{p_k,\rho_k}\sum_kp_kE(\rho_k)
\end{equation}
i.e. the minimum average entanglement of all possible ensemble decompositions of $\rho$, where by $E(.)$ the entropy of entanglement
$E(\rho_{AB})=S(\rho_A)=S(\rho_B)$ is meant, with $S(.)$ the von Neumann entropy.
This measure gives the maximum value of
entanglement that can be achieved by a given unitary $U$ from any possible separable state of purity $\mu$ (purity is not changed
by unitaries). 
Any two-qubit unitary operator can be expressed in Cartan form~\cite{BKJ01} by
\begin{eqnarray}
&U =(L_1\otimes L_2)U_{c}(\alpha_x,\alpha_y,\alpha_z)(L_3\otimes L_4) \\
&U_c(\alpha_x,\alpha_y,\alpha_z)
=\exp(-i\sum_{k=x,y,z}\alpha_k\sigma_k\otimes\sigma_k)
\end{eqnarray}
where $\sigma_k$ are the usual Pauli matrices and $L_i$ are local rotations.
Since entanglement is not increased by local operations, we can assign equivalence classes to every unitary having
the same Cartan kernel $U_c$, that is, the entangling power of a given unitary is a function of its 3-vector $(\alpha_x,\alpha_y,\alpha_z)$ only.
Furthermore, this vector has the following symmetries:\\
a) $U_c(\pi/2+\alpha_x,\alpha_y,\alpha_z)\underset{\rm loc}{=}U_c(\alpha_x,\alpha_y,\alpha_z)$\\
b) $U_c(\pi/4+\alpha_x,\alpha_y,\alpha_z)\underset{\rm loc}{=}U_c(\pi/4-\alpha_x,\alpha_y,\alpha_z)$\\
where we write $\underset{\rm loc}{=}$ for unitaries which are equivalent apart from local rotations~\cite{BKJ01}. Further, these properties
are valid for all angles independently. Using both properties the range for these parameters can be restricted to
\begin{equation}
\pi/4\geq\alpha_x\geq\alpha_y\geq\alpha_z. 
\end{equation}
Also, 
\begin{equation}
U_c(-\alpha_x,\alpha_y,\alpha_z)\underset{\rm loc}{=}U_c(\alpha_x,\alpha_y,\alpha_z)
\end{equation}
because $-\alpha_x\underset{\rm loc}{=}\pi/2-\alpha_x\underset{\rm loc}{=}\pi/4+(\pi/4-\alpha_x)\underset{\rm loc}{=}\pi/4-(\pi/4-\alpha_x)=\alpha_x$.

The final ingredient consists of the knowledge of the mixed states which have maximum entanglement for a given purity, so-called MEMS (maximally entangled mixed states) \cite{M01}.
They are of the form
\begin{equation}
\rho_{\mathrm{ME}}(\gamma, \varphi)=\left(
  \begin{array}{cccc}
    g(\gamma) & 0 & 0 & \frac{\gamma}{2}e^{-i\varphi} \\
    0 & 1-2g(\gamma) & 0 & 0 \\
    0 & 0 & 0 & 0 \\
    \frac{\gamma}{2}e^{i\varphi} & 0 & 0 & g(\gamma)\\
  \end{array}
\right)
\end{equation}
where
\begin{equation}
g(\gamma)=\left\{
  \begin{array}{ll}
    \gamma/2, & \hbox{$\gamma\geq2/3$} \\
    1/3, & \hbox{$\gamma<2/3$}
  \end{array}
\right.
\end{equation}
and the phase $\varphi$ is a rotation of one of the qubits around axis $z$ by an amount $-2\varphi$ which is irrelevant
for entanglement (we include it for later reference). Note that the case $g(\gamma)=\gamma/2$ corresponds to
a rank 2 density matrix, while the case $g(\gamma)=1/3$ corresponds to a rank 3 matrix.
Interestingly, we should realize that $\rho_{\mathrm{ME}}(\gamma,\varphi)$ coincide with the maximally discordant mixed states \cite{qasimi} (see also \cite{FG11})
$\rho_{\mathrm{MD}}(a,b,\varphi)$ when $2/3\leq\gamma\leq1$, which means
that for values of $5/9\leq\mu\leq1$, MEMS and MDMS have the same form.

Finally we can assess one of our main results:

\noindent\textbf{Theorem}: $U_c(\pi/8,\pi/8,\chi), \forall \chi$ is a global two-qubit
entanglement generator, i.e. it has maximum entangling power for all values of
purity. The source separable states that this gate needs to act upon are $\rho^{(R2)}$ and $\rho^{(R3)}$ (denoting that they generate the respective MEMS
of given rank 2 and 3), so that for $\gamma\geq 2/3$ (i.e.
$5/9\leq\mu\leq1$)

\begin{eqnarray}
&&
U_c(\frac{\pi}{8},\frac{\pi}{8},\chi)\rho^{(R2)}U_c^{\dagger}(\frac{\pi}{8},\frac{\pi}{8},\chi)=\rho_{\mathrm{ME}}(\gamma,\pi/2)\\
&& \rho^{(R2)}=\left(
    \begin{array}{cc}
      1-\gamma &  \\
       & \gamma \\
    \end{array}
  \right)\otimes\left(
                  \begin{array}{cc}
                    0 &  \\
                     & 1 \\
                  \end{array}
                \right)
\end{eqnarray}
and for $0\leq\gamma\leq2/3$ (i.e.
$1/3\leq\mu<5/9$), we get
\begin{eqnarray}
U_c(\frac{\pi}{8},\frac{\pi}{8},\chi)\rho^{(R3)}U_c^{\dagger}(\frac{\pi}{8},\frac{\pi}{8},\chi)=\rho_{\mathrm{ME}}(\gamma
, \pi/2)\\
\rho^{(R3)}=\left(
                           \begin{array}{cccc}
                             \frac{1}{3}-\frac{\gamma}{2} &  &  &  \\
                              & \frac{1}{3}&  &  \\
                              &  & 0 &  \\
                              &  &  & \frac{1}{3}+\frac{\gamma}{2} \\
                           \end{array}
                         \right)
\end{eqnarray}
This result can be obtained by direct evaluation and shows that, for general purities, $U_c(\pi/8,\pi/8,\chi)$
reaches the MEMS states, thus becoming a global perfect entangler.

We can also show that when $0\leq\gamma\leq 1/\sqrt3$ (i.e.
$1/3\leq\mu\leq1/2$)
\begin{eqnarray}
&U_c(\pi/4,0,\chi)\rho_{c}U_c^{\dagger}(\pi/4,0,\chi)=\rho_{\mathrm{ME}}(\gamma,
\pi/2)\\
&{\rm with}\ \rho_c=\left(
                           \begin{array}{cccc}
                             \frac{1}{3}+\frac{\gamma}{2} &  &  &  \\
                                & \frac{1}{6} &\frac{i}{6}&  \\
                                & \frac{-i}{6}&\frac{1}{6}&   \\
                               &  &  &\frac{1}{3}-\frac{\gamma}{2} \\
                           \end{array}
                         \right)
\end{eqnarray}
which has zero entanglement for the given $\gamma$ range, thus showing that for this range (coinciding with the range of perfect discording power in \cite{fer13}) it is a perfect
entangler.

We can now ask, is $U_c(\pi/8,\pi/8,\chi)$ the only perfect entangling gate for all purities? The question is rather nontrivial in an algebraic sense,
and we will only give numerical evidence that in fact it is not, though the problem remains open and is left for future investigation.
Our numerical evidence will consist in two complementary approaches: first we produce random states of product form and of classical-classical
form and evaluate their performance for several gates, second, we produce arbitrary gates (restricted to $\alpha_z=0$) plus local rotations
in several axes and apply them to the MEMS states to check whether they can produce separable states for all purities. Finally, we will allow for $\alpha_z\neq0$
to check the validity of the found gates for more general kernels.

\section{Numerical evidence}
\subsection{Performance of specific gates}
\begin{figure}[htbp]
\includegraphics[width=\columnwidth]{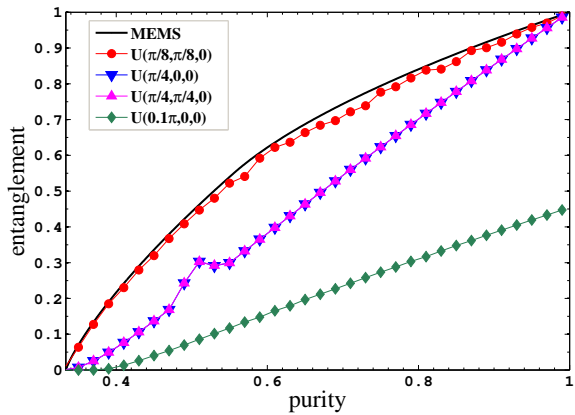}
\caption{(color online) Entangling power versus purity for $(U_c(\pi/8,\pi/8,0)$ (red dots), $U_c(\pi/4,0,0)$ (blue inverted
triangles), $U_c(\pi/4,\pi/4,0)$ (pink triangles), and $U_c(0.1\pi,0,0)$ (dark
green diamonds), based on classical-classical states. We have discretized
$|\alpha_i\rangle$ and $|\beta_j\rangle$ using steps of $0.1\pi$ for each gate
at any purity $\mu$, and generated $1000$ random samples for $p_{i,j}$ for a
pattern of $|\alpha_i\rangle$ and $|\beta_j\rangle$. We see that when $\mu\in[1/3,5/9]$, the
EP$(U_c(\pi/8,\pi/8,0))$ curve overlaps the theoretical curve for MEMS, which
agrees with our analytical result. The loose deviation of the
EP$(U_c(\pi/8,\pi/8,0))$ curve from that of MEMS when $\mu>5/9$ is due to the
small probability of obtaining product states when generating
classical-classical states.}
\label{c-c}
\end{figure}

\begin{figure}[htbp]
\includegraphics[width=\columnwidth]{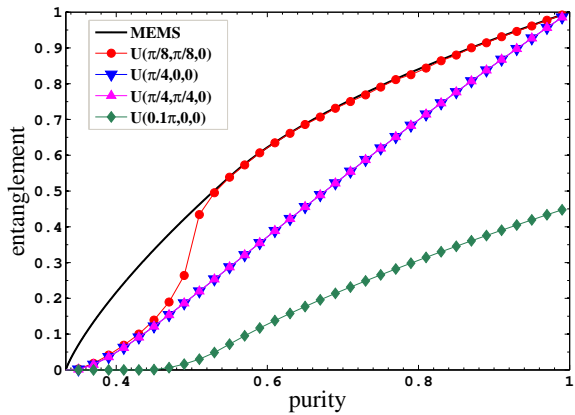}
\caption{(color online) Entangling power versus purity based on product states
$\rho=\rho_{A}\otimes\rho_{B}$. Noting that $\mu=\mu_A\times\mu_B$ we consider
all possible combinations of $\mu_{A}$ and $\mu_{B}$ by steps of $0.01$, and generate more than $1$ million states samples for each purity. 
When $\gamma\leq2/3$, all four gates perform poorly, while the perfect overlapping of the
MEMS and EP$(U_c(\pi/8,\pi/8,0))$ curves when $\mu>5/9$ serves as confirmation
of our analytical result.}
\label{product}
\end{figure}

We begin by considering random states of these two forms
\begin{eqnarray}
&&
\rho^{\mu}_{cc}=\sum_{i,j}p_{i,j}
|\alpha_i\rangle\langle\alpha_i|\otimes|\beta_j\rangle\langle\beta_j|, \mu=\sum_{i,j}p^2_{i,j}\\
&& \rho^{\mu}_{pro}=\rho_{A}\otimes\rho_{B}
\end{eqnarray}
with $|\alpha_i\rangle$, $|\beta_j\rangle$ local orthogonal basis, $p_{i,j}$ probability
distributions at purity $\mu$, and $\rho_A$, $\rho_B$ individual (mixed) states of each qubit. The first states are called classical-classical in the literature
of quantum discord, and are a subset of separable states (separable states do not need $\ket{\alpha_i}$ and $\ket{\beta_i}$ 
to be orthogonal basis).
We scan these states (for details see captions of Figs. \ref{c-c} and \ref{product}) for the 
gates $U(\pi/8,\pi/8,0)$, $U(\pi/4,0,0)$, $U(\pi/4,\pi/4,0)$ and $U(0.1\pi,0,0)$ as shown in the figures.
We make use of the explicit formula of EOF for two-qubits, which is obtained through the well known
concurrence~\cite{H97}.

When a two-qubit system reaches a
level of mixture (here $\mu=1/3$), the entanglement disappears~\cite{TY04}, therefore, the simulation
was only run over the interval $1/3\leq\mu\leq1$. In developing the numerical
evaluation, we considered $\rho_{cc}$ and $\rho_{pro}$ separately. We first
explored the performance of $\textrm{{EP}}_{\mu}(U_c)$ acting on $\rho_{cc}$. In
Fig.~\ref{c-c}, we can see that $U_c(\pi/8, \pi/8, 0)$ is a perfect entangler at
low purity intervals ($1/3\leq\mu\leq5/9$) since the
$\textrm{{EP}}_{\mu}(U_c)$ curve overlaps with the EOF of the MEMS. However,
when $\mu$ enters the $[5/9,1]$ interval, the $\textrm{{EP}}_{\mu}(U_c(\pi/8,
\pi/8, 0))$ curve loosely deviates from the curve for the theoretical maximum.
If we let $\textrm{{EP}}_
{\mu}(U_c)$ act on $\rho_{pro}$, however, we see in Fig.~\ref{product} that
$U_c(\pi/8, \pi/8, 0)$ possesses excellent entangling power when $\mu$ enters
$[5/9,1]$, since the $\textrm{{EP}}_{\mu}(U_c(\pi/8, \pi/8, 0))$ curve closely
matches the theoretical line. 

In a numerical evaluation with classical-classical states, the
possibility of generating a direct product state is small, so it is easy to understand
that in Fig.~\ref{c-c}, the gate $U(\pi/8,\pi/8,0)$ is not able to fully reach 
MEMS states for $2/3\leq\gamma\leq1$, since the ideal source state is $\rho^{(R2)}$, which is a product state. In
the same way, in Fig.~\ref{product} such gate performs very poorly for $0\leq\gamma\leq2/3$ because a product state
cannot reproduce the ideal source (a classical-classical state) $\rho^{(R3)}$.

Based on both numerical simulations, however, we can
say that $U_c(\pi/8,\pi/8,0)$ has maximal entangling power and serves as perfect
entangler at any purity $\mu$ when acting on zero entanglement states, which
supports our analytical proof.


We note that $U_c(\pi/4,\pi/4,0)$ and
$U_c(\pi/4,0,0)$(kernel of the CNOT) have the same entangling power but it is much smaller than that of
$U_c(\pi/8,\pi/8,0)$. Finally, the gate $U_c(0.1\pi,0,0)$ performs very badly.

\subsection{Are there more global perfect entanglers?}

\begin{figure}[h!]
\includegraphics[width=0.8\columnwidth]{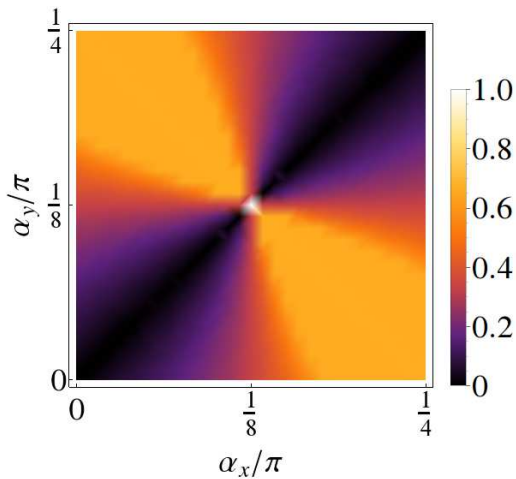}
\caption{We evaluate $U_c(\alpha_x,\alpha_y,0)$ with an additional local rotation in $z$-axis (which is scanned, see main text)
acting on the MEMS. For each value of
$\{\alpha_x,\alpha_y \}$ we calculate for how many values of $\gamma$ (in proportion) such gate can reach a separable state.
We see that only $\alpha_x=\pi/8$, $\alpha_y=\pi/8$ reaches all purities for the full range of $\gamma$, while many of the gates (see orange plateau) can perform
as perfect entanglers only for rank 3 states (not shown).}
\label{fig3}
\end{figure}

We inquire next about the existence of other global perfect entanglers
through a complementary approach, namely: we take the MEMS states, apply
a given gate (rather its inverse $U_c^\dagger$) and check whether they can achieve a separable state for
any $\gamma$. If they cannot, they are not global, and we store the proportion
of the $\gamma$ range in which the gate is a perfect entangler. This
problem is involved both algebraically and numerically, since we need
to check the condition $E_F[U_c^\dagger(L_A^\dagger\otimes L_B^\dagger)\rho_{\rm ME}(L_A\otimes L_B)U_c]=0$
considering all possible local rotations $L_A$ and $L_B$ (each of them
parametrized by two independent angles).
We restrict ourselves first 
to local rotations in the $z$-axis, the result is shown in Fig.~\ref{fig3} 
for gates $U_c(\alpha_x,\alpha_y,0)$. 
It is observed that only $\alpha_x=\pi/8$, $\alpha_y=\pi/8$ is a global perfect entangler, 
while many of the gates (see orange plateau) can perform as perfect entanglers
only for $2/3$ of the range. The value $2/3$
we observe in this figure can be understood if we
do the same analysis but only storing the range $\gamma$ restricted by ranks (not shown). 
That is, we separate the ranges $2/3\leq\gamma\leq 1$ (rank 2)
and $0\leq\gamma< 2/3$ (rank 3) and observe that the orange plateau in Fig.~\ref{fig3} is a global
perfect entangler for rank 3, while for rank 2 only $(\alpha_x=\pi/8,\alpha_y=\pi/8)$ is global.
The orange plateau consists of gates in the neighborhood of $\alpha_x+\alpha_y=\pi/4$ such as $U_c(\pi/4,0,0)$. We
also stress the notable fact that gates $U_c(\chi,\chi,0)$ are never perfect entanglers (except when $\chi=\pi/8$).

Performing the same analysis but instead with local rotations in the $x$-axis (not shown), it turns out
that gates $U_c(\chi,\pi/4,0)$ are also global perfect entanglers $\forall\chi$.
When local rotations are done around $y$-axis we observe that $U_c(\pi/4,\chi,0)$ is global perfect entangler, as
it should be, since the labels $x,y,z$ are arbitrary. Furthermore, we checked other combinations of local
rotations, namely about local axes $(x,y)$, $(x,z)$ and $(y,z)$, and only corroborated the latter gates.

With this complementary approach we have discovered a new global perfect entangler gate, $U_c(\pi/4,\chi,0)$. We can now
ask: is this gate (and the $U_c(\pi/8,\pi/8,0)$ one) performing as well if we pick a different $\alpha_z$?.
To answer this question, we assign the values $\alpha_z=\pi/16,\pi/8,\pi/6,\pi/4$ and evaluate numerically again with local rotations. This
analysis extracts only the gates $U_c(\pi/4,0,\alpha_z)$, $U_c(0,\pi/4,\alpha_z)$ (with the former values of $\alpha_z$) as
global perfect entanglers. That is, the only combinations which are global perfect entanglers are $U_c(\pi/4,0,\chi)$ and $U_c(\pi/8,\pi/8,\chi)$
with all possible permutations of the indices.

\begin{figure}[h!]
\includegraphics[width=0.8\columnwidth]{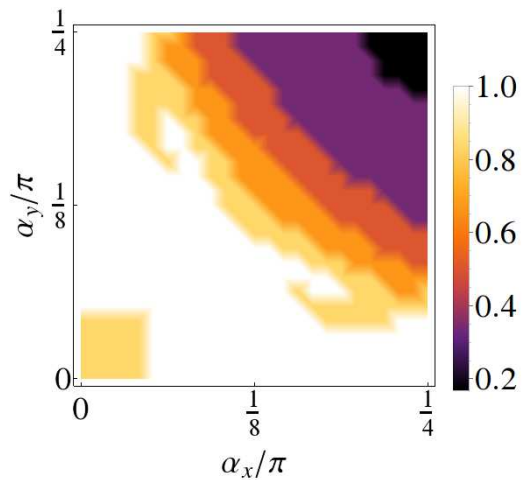}
\caption{Evaluation of $U_c(\alpha_x,\alpha_y,\pi/6)$ with additional local rotations in $(x,y)$-axis (for qubit A, B respectively), only for rank 3
states. A large region of gates exists with perfect entangling power for this rank. }
\label{fig4}
\end{figure}

Surprisingly, however, we find a region of gates $U_c(\alpha_x,\alpha_y,\pi/6)$ which, up to numerical precision 
of parameter scan, seem to be perfect entanglers for rank 3; they are shown in figure~\ref{fig4}. 
These gates, though not global, emphasize the difficulty of finding general perfect entangling
gates when all parameters in the problem are considered. Therefore, the question of exhausting all possible perfect entangler
gates, global or not, remains an open problem for future investigation.


\section{DISCUSSION}

We have investigated the problem of finding the maximum amount of entanglement
that can be produced by a two-qubit unitary operator from any separable state.
The problem is reduced to finding such entangling power for the equivalence classes
dictated by the 3 parameters $\alpha_x,\alpha_y,\alpha_z$, due to the fact that
entanglement is unchanged by local rotations and because unitary operators can be
split by Cartan decomposition into such rotations and a nonlocal (two-qubit)
kernel. We were able to show analytically that all gates with kernel $U_c(\pi/8,\pi/8,\chi)$
are global perfect entanglers, i.e. one can always find a separable state for each
possible purity such that this gate promotes the latter to a maximally entangled
mixed state (MEMS) of the given purity. It is hard to fail noticing that
this gate was also shown to be a global perfect discorder (i.e. produces maximum
quantum discord \cite{discord}), which is a curious fact which deserves further investigation.

We performed two complementary numerical
investigations to complement our analytical results: first, we produced many random states of
product and classical-classical form, thereby testing several gates. We found
that the gate above is, within these source families of separable states which do not
exhaust all possible separable states, a global perfect entangler, while other gates
(like $U_c(\pi/4,0,0)$ and $U_c(\pi/4,\pi/4,0)$ did not a priori seem to be perfect entanglers
for all purities, but only for a given range.
Second, we went the other way around: start with the MEMS states, perform local rotations
on them and apply all possible gates $U_c(\alpha_x,\alpha_y,0)$, then check whether a
separable state can be obtained. With this procedure, always limited by
computational difficulty due to many parameters, we found that the gates with kernel
$U_c(\pi/4,\chi,0)$ are also global perfect entanglers. It should be stressed that 
this finding was not obtained with the numerical evaluation of figures~\ref{c-c},\ref{product},
meaning that MEMS cannot be reached neither from product nor classical-classical states.
This might be the reason why discording power, as defined in \cite{fer13}, cannot be
reached for high purities with such gate, but it does for entangling power (considering
that discord and entanglement have the same value for this range of MEMS \cite{plastina}).
In addition, we found several families of gates that, though not global, are perfect entanglers
reaching the rank 3 MEMS, such as the neighbors of $U_c(\chi,\pi/4-\chi,0)$ (see fig.~\ref{fig3})
and a whole region of gates of the form $U_c(\alpha_x,\alpha_y,\pi/6)$ (see fig.~\ref{fig4}).

Although the problem of finding all global perfect entangler gates is very hard both
analytically and numerically, we believe we have made a step forward in finding
some of them, which might be of help in devising experimental setups which can
take advantage of mixed states as a source for producing entanglement. At the same
time, it can help understand the theoretical difference between several quantumness measures
as entanglement or discord, and help gain insight into the different hierarchies
of gates in terms of production of quantum advantage for quantum computation.

Our work is thus a first attempt to quantify the power of two-qubit gates in
generating entanglement in mixed states, thus providing a way to specifically
analyze entanglers for general purity values. The power of
$U_c(\pi/8,\pi/8,\chi)$ gates to produce maximum entanglement along with discord can provide some unique
experimental utilization of such gates in quantum computation and other
areas.

\section{Acknowledgement}

This work was supported by the National Basic Research Program of China (Grant
No.~ 2011CB921200), the CAS, the National Natural Science Foundation of China
(Grant Nos. 11274289 and 11105135 ), the Fundamental Research Funds for the
Central Universities (Nos. WK2470000011, WK2470000004, WK2470000006, and
WJ2470000007). ZG and HH acknowledge support from the Fund for Fostering Talents
in Basic Science of the National Natural Science Foundation of China (No.~
J1103207). FG acknowledges MICINN, MINECO, CSIC, EU
commission and FEDER funding under grants FIS2007-
60327 (FISICOS), FIS2011-23526 (TIQS), post-doctoral JAE
program (ESF) and COST Action MP1209.

\end{document}